# МЕТОД СИНТЕЗА НАЛОЖЕННОЙ СЕТИ MPLS С ИСПОЛЬЗОВАНИЕМ МАТЕМАТИЧЕСКОЙ МОДЕЛИ МНОГОСЛОЙНОГО ГРАФА

Агеев Д. В., Хайдара Абдалла, Старкова Е.В.
*Харьковский национальный университет радиоэлектроники*
г. Харьков, пр. Ленина,14, 61166, Украина
тел.: +380(57)702-13-20, e-mail: dm@ageyev.in.ua

*Аннотация* — В работе предложено в качестве математической модели при решении задачи проектирование сети MPLS использовать многослойный граф. Применение данной модели позволяет проводить проектирование мультисервисных телекоммуникационных систем одновременно на нескольких их уровнях и свести решаемую задачу к задаче поиска подграфа минимального веса. Проведено исследование эффективности предложенного метода, которое показало, что предложенный метод позволяет получить структуру сети на 10-16 % меньшей стоимости.

## I. Введение

Быстрое развитие информационных технологий требует развертывания высокоскоростных сетей, обеспечивающих возможность передачи трафика с гарантированным качеством обслуживания. Одной из таких технологий является MPLS. Стандарт MPLS не включает спецификации нижних уровней модели ВОС и предусматривает развертывание данных сетей как наложенных.

Методы синтеза структуры сети MPLS предлагались и ранее, так например в работе [5] рассмотрен метод синтеза телекоммуникационных сетей MPLS с учетом эффекта статистического мультиплексирования, который позволяет также решать задачу определения мест расположения узлов с коммутацией пакетов и оптимизации виртуальной топологии сети. Ряд других работ [6, 7] также посвящены задаче определения мест расположения узлов сети с функциями MPLS.

При синтезе наложенных сетей необходимо определять топологию сети на каждом из ее уровней. При этом подход базирующийся на последовательном синтезе каждого уровня, с использованием результатов синтеза одного уровня, как исходных данных для синтеза соседнего, не дает оптимального решения задачи в целом. Для устранения данного недостатка в работах [10, 11] предложено для моделирования наложенных сетей использовать многослойный граф, который позволяет учитывать технологическую многоуровневую природу современных телекоммуникационных систем.

В данной работе предложен метод совместного решения задач определения мест размещения узлов с функциями MPLS, выбора пропускных способностей каналов связи и производительности узлов сети. Решение задачи базируется на применении математической модели многослойного графа.

## II. Постановка задачи

Рассматриваемая в докладе сеть MPLS является многоуровневой и содержит два уровня: транспортная сеть и сеть MPLS.

В сети MPLS связи между узлами обеспечиваются транспортной сетью. В качестве транспортной сети могут использоваться сети SDH (Synchronous Digital Hierarchy) или WDM (Wavelength Division Multiplexing) в зависимости от требуемой пропускной способности. В такой сетевой архитектуре, некоторые или все узлы сети поддерживают не только технологии транспортной сети, но технологию MPLS.

Таким образом, при проектировании данного класса сетей необходимо определять топологию обоих сетей: топологию транспортной сети и топологию сети MPLS. Это означает, что мы должны определить:
- какие узлы транспортной сети должны поддерживать функциональность MPLS;
- каким образом должны быть связаны узлы LSR через транспортную сеть;
- какую величину пропускной способности должны иметь соединения между LSR.

В общем случае, решением задачи является установка оборудования MPLS в некоторой части узлов транспортной сети, так чтоб обеспечить компромисс между затратами на оборудование узлов сети функциями MPLS и расходами, связанными с неполным использованием пропускной способности каналов связи.

При решении задачи синтеза необходимо учитывать эффект агрегирования трафика на дополнительном логическом уровне и многоуровневую природу современных телекоммуникационных систем.

## III. Метод решения задачи

Согласно общей методике решения задачи синтеза мультисервисных телекоммуникационных систем с использованием многослойных графов [5], мы должны синтезировать начальный избыточный МСГ. Для этого мы должны выполнить следующие шаги:
- выделить в синтезируемой сети отдельные наложенные слои;
- каждый слой описать графом, описывающим связи на каждом из слоев;
- задать ребра, образующие связи между слоями;
- присвоить ребрам и вершинам МСГ набор величин, характеризующих значения параметров соответствующих элементов и связей моделируемой системы.

Из анализа предметной постановки задачи можно выделить следующие слои:
- нижним слоем многослойного графа является слой, описывающий топологию транспортной сети;
- второй слой, располагающийся выше первого, соответствует уровню MPLS - сети;
- вышележащие слои соответствуют потокам, передаваемым через проектируемую сеть.

Синтезированный многослойный граф используется для решения задачи синтеза сети MPLS.

Решение задачи сводится к нахождению многослойного подграфа минимального веса, обеспечивающего пропускание информационных потоков, с учетом требований к структуре многослойного графа [5] и потокам, протекающим по его ребрам [6], при выполнении ограничений на пропускную способность ребер многослойного графа.

## IV. Исследование эффективности предложенного метода

Экспериментальное исследование предложенного метода заключается в сравнительном его анализе с ме-тодом, предусматривающим установку оборудования LSR в каждом из узлов базовой сети.

При применении альтернативного метода (без учета многослойной структуры), суть решаемой задачи сводилась к задаче оптимального распределения потока, который обеспечивал бы минимальное количество используемых оптических каналов связи.

Результаты экспериментального исследования представим в виде таблицы (табл. 1).

Табл. 1. Результаты сравнительного анализа
Tabl. 1. Comparative analysis results

| Кол-во узлов | Номер варианта набора исходных данных | | | | | | | |
|---|---|---|---|---|---|---|---|---|
| | 1 | | 2 | | 3 | | 4 | |
| 20 | 671 | 717 | 1086 | 1183 | 502 | 557 | 707 | 749 |
| 25 | 801 | 865 | 1950 | 2106 | 699 | 768 | 1518 | 1654 |
| 30 | 1608 | 1752 | 3304 | 3634 | 1153 | 1256 | 2470 | 2717 |
| 35 | 2858 | 3143 | 6861 | 7752 | 1901 | 2110 | 2836 | 3034 |
| 40 | 6081 | 6749 | 10138 | 11050 | 2439 | 2731 | 4550 | 5187 |
| 45 | 7276 | 8149 | 21138 | 23463 | 4690 | 5252 | 8650 | 9342 |
| 50 | 14296 | 16011 | 21359 | 24562 | 7921 | 8713 | 13570 | 14927 |

Анализ приведенных результатов показал, что использование метода, базирующегося на применении многослойного графа, позволяет получить структуру сети, которая по своим стоимостным показателям меньше стоимости структуры, полученной альтернативным методом в среднем на 10 – 16 %.

## V. Заключение

В докладе приведена постановка и решена задача проектирования сети MPLS согласно критерию минимум стоимости. Для решения поставленной задачи целесообразно в качестве математической модели использовать многослойный граф, что позволяет учитывать многоуровневую природу современных телекоммуникационных систем, образуемую наложенными сетями и осуществлять решение задачи с определение структуры сети MPLS одновременно на двух ее уровнях (уровень сети MPLS и уровень транспортной сети).

В работе решение задачи проектирования мультисервисной телекоммуникационной системы сводиться к задаче нахождения многослойного подграфа минимального веса, с учетом ограничений на пропускную способность ребер графа.

Исследование эффективности предложенного метода показало, что он позволяет получить структуру сети с меньшей стоимостью на 10 – 16 %.

Предложенный метод решения задачи может быть использован на практике, при проектирования сетей NGN на уровне транспортной.

## VI. Список литературы

# MPLS OVERLAY NETWORK SINTHESIS METHOD WITH MULTILAYER GRAPH USAGE

Ageyev D. V., Haidara Abdalla, Starkova O. V.
*Kharkov national university of radioelectronics*
*Lenina av., 14, Kharkov, 61166, Ukraine*
Ph.: +380(57)7021320, e-mail: dm@ageyev.in.ua

*Abstract* —It is suggested to use multi-layer graphs as a mathematical model in the design of MPLS networks. The application of this model makes it possible to design multi-service telecommunication systems simultaneously at several levels and to reduce the problem to the search of the minimum weight graph.

## I. Introduction

Modern telecommunication systems are multi-layer by their structure. Taking into the account of the modern multi-level systems requires the development of new mathematical models, which would allow to adequately describe the existing physical and logical connections between the elements of the system on its different levels, different types of hierarchies, and to effectively solve the problems of design.

The paper formulates the problem of synthesis of structure of MPLS network layered with the transport SDH network or WDM and suggests the method of its solving. The solving is based on the application of mathematical model of multi-layer graph.

## II, III, IV. Main Part

The MPLS network considered in the thesis is multilevel and has two levels: transport network and MPLS network. Thus, during the planning of such networks it is necessary to determine the topology of both the networks: that of transport one and that of MPLS one. This means that one needs to determine:
- what nodes of the transport network should support the MPLS functionality;
- in what way the LSR nodes should be connected via the transport network;
- what should be the bandwidth between LSR links.

According to the general method for solving the problem of synthesis of multiservice telecommunication systems with the usage of multi-layer graphs, we have to synthesize the initial redundant $MLG = (\Gamma, V, E)$.

The solving of the task above can be reduced to the finding of the multi-layer minimum weight subgraph $MLG' \subset MLG$ that provides the transfer of information flows with the consideration of the requirements to the structure of the multilayer graph [5] and the flows on its edges [6] observed at the applying of constraints to the bandwidth of the edges of the multi-layer graph.

## V. Conclusion

The paper reduces the problem of design of multiservice telecommunication system with the transferred multicast flows to the problem of finding a multilayer minimum weight subgraph with the consideration of constraints to the graph edges bandwidth.

It is shown that application of given method provide to reduce MPLS network cost to 10 – 16 %.